\documentclass[prl,preprintnumbers,twocolumn,showpacs]{revtex4}
\usepackage{amsfonts,mathrsfs}
\usepackage{amsmath,rotating,calc,graphics,color}

\begin{document}

\title{\Large{\bf Correction to the Running of Gauge Couplings \\
out of Extra Dimensional Gravity } }
\author{Feng Wu$^{1}$}
\email[Electronic address: ]{fengwu@ncu.edu.cn}
\author{Ming Zhong$^{2}$}
\email[Electronic address: ]{zhongm@nudt.edu.cn}
\affiliation{${}^{1}$ Department of Physics, Nanchang University,
330031, China}
 \affiliation{${}^{2}$ Department of Physics, National
University of Defense Technology, Hunan 410073, China}

\begin{abstract}
We study the gravitational correction to the gauge couplings in the
extra dimension model where the gravity propagates in the
(4+n)-dimensional bulk. We show by explicit calculation in the
background field method that the one-loop correction coming from
graviton Kaluza-Klein states is nontrivial and tends to make the
theory anti-asymptotically free. For the theory characterized with
asymptotic freedom, the correction will induce a nontrivial
UV-stable fixed point. This may lead to an interesting possibility
that the non-abelian gauge coupling constants $g_2$ and $g_3$ in the
SM and that of the gravity $g_{\kappa}$ unify at the fixed point in
the limit of high energy scale.
\end{abstract}
 \maketitle 
\date{\today}
General relativity is not renormalizable after quantization
\cite{DeWitt,tHooft}. This is one of the reasons that general
relativity was considered to be incompatible to quantum mechanics. A
modern point of view to the non-renormalizable theory is that it
might be sensible and reliable predictions could still be made from
it within the framework of effective field theories. From the value
of the only dimensional coupling constant $G$, Newton's constant, in
Hilbert-Einstein Lagrangian, one can see that gravitational effects
are tiny at energies $\mu \ll M_{pl} \sim 10^{19} GeV/c^2$. It makes
sense to treat general relativity as a low energy effective field
theory of some unknown fundamental theory and consider its quantum
effects \cite{Donoghue}. The effects due to non-renormalizable terms
are suppressed by inverse powers of $M_{pl}$, the mass scale of new
physics.

The most promising candidate for such new physics is the string
theory. A consistent string theory predicts a definite number of
extra dimensions. To make contact between string theory and our
experienced four-dimensional spacetime, the most straightforward
possibility is that the extra dimensions are compactified on an
internal manifold whose size is sufficiently small to escape being
detected at present.

The field theory with compactified extra dimensions is another
example of the non-renormalizability different from the gravitation.
When a renormalizable theory in four-dimensional spacetime is
generalized to (4+n)-dimensions and reduced to the four-dimensional
world, the effective lagrangian contains infinite Kaluza-Klein
series, with every operator having mass dimension less than five.
Considering that the heavy Kaluza-Klein states are decoupled from
the low energy physics of phenomenological interest, one can
truncate the Kaluza-Klein tower at some energy scale $\mu$. The
truncated theory is renormalizable.

An interesting thing concerned with these two types of effective
theory is the quantum corrections to the gauge couplings. In
\cite{Robinson}, Robinson and Wilczek showed that gravitational
corrections will cause gauge couplings to run quadratically.
Although the effect is tiny in the regions where perturbative
calculations are reliable, it's of theoretical interest on its own.
Pietrykowski \cite{Pietrykowski} redid the calculations and found
that the result in \cite{Robinson} is gauge dependent. Up to
one-loop order the gravitational contributions to the $\beta$
function are zero in the dimensional regularization (DR) scheme.
Later on, Pietrykowski's result was verified by
\cite{wu1,toms,ebert}. More recently, it was argued that the
gravitational contribution to the $\beta$ function is nontrivial in
a new regularization scheme \cite{tang}.

The power law running exist in the theory of n extra dimensions. The
one-loop correction to the gauge couplings from a tower of gauge
boson Kaluza-Klein states is of order $\mu^{n}$ \cite{Dienes}. It
drives the theories to run fast so that the gauge coupling
unification can be significantly lowered to TeV scale in large extra
dimension models.

In this work, we will calculate in DR scheme the one-loop correction
to the gauge couplings from the extra dimensional gravity where the
gravitons propagate in (4+n) dimensional spacetime while the gauge
and matter fields live in the normal four dimensional world
\cite{Arkani-Hamed}. We will show, by explicit calculations in the
background field method, that the one-loop correction coming from
graviton Kaluza-Klein states is nontrivial. It is universal for all
gauge couplings and tends to make the theory anti-asymptotically
free. Therefore for the theory characterized with asymptotic
freedom, the correction will induce a nontrivial UV-stable fixed
point. This may lead to an interesting possibility that the
non-abelian gauge coupling constants $g_2$ and $g_3$ in the SM and
that of the gravity $g_{\kappa}$ unify at the fixed point in the
limit of some high energy scale, if the numbers of the active
Kaluza-Klein particles are properly selected.

We start with the Einstein-Yang-Mills theory. The extra dimensions
are compactified on a torus $T^n$. The compactification scales of
the extra n-dimensional spaces $y_i$ are all assumed to be roughly
equal to $R$, and they need not to be large extra dimensions in our
present consideration. The action of the theory has the form:
\begin{equation}
S= - \int d^4x d^ny ({\cal L}_{HE}+{\cal L}_{YM}), \label{action}
\end{equation}
where the Hilbert-Einstein and Yang-Mills lagrangians are
\begin{eqnarray}
{\cal L}_{HE}&=&{1\over \hat{\kappa}^2} \sqrt{ (-1)^{3+n}|\hat{g}^{(4+n)}|}\hat{R}, \nonumber\\
{\cal L}_{YM}&=&{1\over4}\sqrt{ - |\hat{g}^{(4)}|} \hat{g}^{\mu
\lambda} \hat{g}^{\nu \rho} F^a_{\mu \nu} F^a_{\lambda
\rho}\delta^{(n)}(y),
\end{eqnarray}
with $\hat{\kappa}^2 \equiv \frac{16 \pi}{M_{pl(4+n)}^{2+n}}$, and
the relation $\kappa^2R^n=\hat{\kappa}^2$. The symbols $\hat{R}$,
$\hat{g}^{\hat{\mu} \hat{\nu}}$ and $M_{pl(4+n)}$ are the Ricci
scalar, metric tensor and \emph{Planck} mass in (4+n) bulk. The
index $\hat{\mu}$ extends over the full (4+n) dimensions and $\mu$
over the (3+1) dimensions. The gauge coupling $g$ is included in the
field strength
\begin{equation}
F^a_{\mu
\nu}=\mathcal{D}_{\mu}A^a_{\nu}-\mathcal{D}_{\nu}A^a_{\mu}+gf^{abc}A^b_{\mu}A^c_{\nu},
\end{equation}
where $\mathcal {D}_{\mu}$ is the spacetime covariant derivative
operator.

In the following we use the background field method \cite{Abbott}
and choose the background spacetime to be flat. Since the theory
respects Poincar$\acute{e}$ invariance in flat spacetime, the gauge
coupling constant is universal everywhere and the following
discussion is without ambiguity. As is well known, using the
background field method one is able to quantize a gauge field theory
without losing the explicit gauge invariance. In this framework, the
$\beta$ function of the gauge coupling constant can be calculated
from the renormalization constant of the background gauge field.
This greatly simplifies the calculations, especially in the
non-abelian and higher loop cases.

For later convenience, we illustrate the procedure for calculating
the $\beta$ function of non-abelian gauge theories in curved
spacetime. To renormalize the divergence, the bare and renormalied
quantities are related by
\begin{eqnarray}
\bar{A}_{0\mu}^a &=&Z_A^{\frac{1}{2}}\bar{A}_{\mu}^a ,\\
g_0&=&Z_gg ,\label{g0}\\
\bar{F}^a_{0\mu\nu}&=&Z_A^{\frac{1}{2}}(\partial_{\mu}\bar{A}_{\nu}^a-\partial_{\nu}\bar{A}_{\mu}^a+Z_A^{\frac{1}{2}}Z_ggf^{abc}\bar{A}^b_{\mu}\bar{A}^c_{\nu})
,
\end{eqnarray}
where $\bar{A}^a$ are background gauge fields. Because explicit
gauge invariance is retained, the gauge covariant form of the
strength tensor implies the relation
\begin{eqnarray} Z_g=Z_A^{-\frac{1}{2}} . \label{zg}\end{eqnarray}

In DR scheme, we perform all loop momentum integrals in
$d=4-\varepsilon$ dimensions. The bare coupling constants are not
dimensionless. To use a dimensionless renormalized coupling
constant, we introduce an arbitrary mass parameter $\mu$ and replace
the Eq.(\ref{g0}) by \begin{eqnarray} g_0=Z_g\mu^{\varepsilon}g .
\label{g01}\end{eqnarray} The renormalized constant $g$ are
functions of $\mu$, while $g_0$ is independent of $\mu$. By
differentiating both sides of Eq.(\ref{g01}), we find the $\beta$
function
\begin{eqnarray}
\beta\equiv \mu\frac{\partial g}{\partial \mu}=-\varepsilon
g-g\mu\frac{\partial \ln Z_g}{\partial \mu}.\label{beta}
\end{eqnarray} Using the chain rule
\begin{eqnarray}
\mu\frac{\partial}{\partial \mu}=\mu\frac{\partial g}{\partial
\mu}\frac{\partial}{\partial g}+ \mu\frac{\partial \kappa}{\partial
\mu}\frac{\partial}{\partial \kappa},
\end{eqnarray}
$\beta$ function can be written in a simple form \begin{eqnarray}
\beta=-\varepsilon g-g\beta\frac{\partial \ln Z_g}{\partial
g}-g\beta_{\kappa}\frac{\partial \ln Z_g}{\partial
\kappa}.\end{eqnarray} The definition of the $\beta$ function for
gravity $\beta_{\kappa}$ is the same as that of the gauge sector. It
has the form
\begin{eqnarray}
\beta_{\kappa}=-\varepsilon \kappa-\kappa\mu\frac{\partial \ln
Z_{\kappa}}{\partial \mu}.\label{betak}
\end{eqnarray}
From Eqs.(\ref{zg}), (\ref{beta}) and (\ref{betak}) and taking the
limit $\varepsilon\rightarrow 0$, one can finally have
\begin{eqnarray}
\beta=-\frac{1}{2}\varepsilon g^2\frac{\partial \ln Z_{A}}{\partial
g}-\frac{1}{2}\varepsilon g\kappa\frac{\partial \ln Z_{A}}{\partial
\kappa}, \label{beta1}
\end{eqnarray}
where the first term purely comes from the gauge and matter sector,
and the second term results from the gravitational correction. It
shows that the $\beta$ function can be determined from the
coefficient of the $1/\varepsilon$ term in the gauge field two-point
Green's function.

 Now we continue to calculate the correction to the two-point
Green's function of the gauge field from $n$-extra dimensional
gravity. Fields $\hat{g}_{\hat{\mu}\hat{\nu}}(x,y)$ and $A^a(x)$ can
be written as sums of background fields
$(\hat{\eta}_{\hat{\mu}\hat{\nu}}, \bar{A}^a(x))$ and quantum
fluctuations $( \hat{h}_{\hat{\mu}\hat{\nu}}, a^a(x))$:
\begin{eqnarray}
&& \hat{g}_{\hat{\mu}\hat{\nu}}(x,y)=\hat{\eta}_{\hat{\mu}\hat{\nu}}
+ \hat{\kappa} \hat{h}_{\hat{\mu}\hat{\nu}}(x,y),\nonumber\\
&& A^a(x) = \bar{A}^a(x) + a^a(x). \label{fields}
\end{eqnarray}
Here $\hat{\eta}_{\hat{\mu}\hat{\nu}}$ is the Minkowski metric of
the bulk.

To perform the Kaluza-Klein reduction to four-dimensional spacetime,
we parameterize the field $\hat{h}_{\hat{\mu}\hat{\nu}}$ as
\begin{equation}
\hat{h}_{\hat{\mu}\hat{\nu}}=V^{-{1\over2}}_n\left(
  \begin{array}{cc}
    h_{\mu\nu}+\eta_{\mu\nu}\phi & A_{\mu j} \\
    A_{i\nu} & 2\phi_{ij} \\
  \end{array}
\right),
 \end{equation}
where $V_n=R^n$ is the volume of the $n$-dimensional compactified
torus $T^n$, $\phi=\sum_i\phi_{ii}$, the subscript $\mu,\nu=0,1,2,3$
and $i,j=4,5,...,3+n$. The fields $h_{\mu\nu}$, $A_{\mu i}$ and
$\phi_{ij}$ are Lorentz tensor, vector and scalar respectively. They
have the following mode expansions:
\begin{eqnarray}
h_{\mu\nu}(x,y)&=&\sum_{\vec{n}}h_{\mu\nu}^{\vec{n}}(x)exp(i\frac{2\pi\vec{n}\cdot\vec{y}}{R}),\\
A_{\mu i}(x,y)&=&\sum_{\vec{n}}A_{\mu i}^{\vec{n}}(x)exp(i\frac{2\pi\vec{n}\cdot\vec{y}}{R}),\\
\phi_{ij}(x,y)&=&\sum_{\vec{n}}\phi_{ij}^{\vec{n}}(x)exp(i\frac{2\pi\vec{n}\cdot\vec{y}}{R}),
\end{eqnarray}
with $\vec{n}={(n_1,n_2,...,n_n)}$.

To continue, one has to fix gauges. Adding a special de Donder gauge
fixing term
\begin{equation}-{1 \over 2}(
\partial_{\hat{\rho}} \hat{h}^{\hat{\rho} \hat{\mu}} \partial^{\hat{\sigma}} \hat{h}_{\hat{\sigma} \hat{\mu}}- \partial_{\hat{\rho}} \hat{h}^{\hat{\rho}
\hat{\mu}}
\partial_{\hat{\mu}} \hat{h}+\frac{1}{4}\partial_{\hat{\mu}} \hat{h}\partial^{\hat{\mu}} \hat{h})\end{equation}
for the graviton field and the Lorentz gauge fixing term ${-1\over 2
\xi} (\partial_{\mu} a^a_{\nu} )^2$ for the photon field to the
action, the propagators for the massive Kaluza-Klein states
$h^{\vec{n}}_{\mu\nu}$, $A^{\vec{n}}_{i\mu}$, $\phi^{\vec{n}}_{ij}$
and the quantum gauge field in momentum space are:
\begin{eqnarray}
&&\triangle^h_{\vec{n}\mu\nu,\vec{m}\rho\sigma}(k)=-i\frac{\delta_{\vec{n},-\vec{m}}(\eta_{\mu\rho}\eta_{\nu\sigma}+\eta_{\mu\sigma}\eta_{\nu\rho}-\eta_{\mu\nu}\eta_{\rho\sigma})}
{k^2-m_{\vec{n}}^2+i\epsilon}\nonumber\\
&&\triangle^A_{\vec{n}i\mu,\vec{m}j\nu}(k)=-i\frac{\delta_{\vec{n},-\vec{m}}\delta_{ij}\eta_{\mu\nu}}{k^2-m_{\vec{n}}^2+i\epsilon}\nonumber\\
&&\triangle^\phi_{\vec{n}ij,\vec{m}kl}(k)=-i\frac{\delta_{\vec{n},-\vec{m}}[\frac{1}{4}(\delta_{ik}\delta_{jl}+\delta_{il}\delta_{jk})-\frac{1}{4+2n}\delta_{ij}\delta_{kl}]}
{k^2-m_{\vec{n}}^2+i\epsilon}\nonumber\\
&&\triangle^a_{a\mu,b\nu}(k)={-i\delta_{ab} \over k^2 + i \epsilon}
[ \eta_{\mu\nu} -(1- \xi) { k_{\mu} k_{\nu} \over k^2}],
\end{eqnarray}
where $m^2_{\vec{n}}\equiv \frac{4\pi^2n_in_i}{R^2}$ is the mass of
Kaluza-Klein graviton at n-th order excitations. Obviously the
spin-2 state $h^{\vec{n}}_{\mu\nu}$ is not a physical state. This
can be seen explicitly from the numerator of its propagator. The
physical massive spin-2 state with right polarization tensor is
constructed from $h^{\vec{n}}_{\mu\nu}$, $A^{\vec{n}}_{i\mu}$ and
$\phi^{\vec{n}}_{ij}$. However, the rearrangement of the physical
spin-2, $(n-1)$ spin-1, and $n(n-1)/2$ spin-0 states into
$h^{\vec{n}}_{\mu\nu}$, $A^{\vec{n}}_{i\mu}$ and
$\phi^{\vec{n}}_{ij}$ states shown above greatly simplifies our
calculations.

\begin{figure}
\begin{center}
\includegraphics[width=8cm,clip=true,keepaspectratio=true]{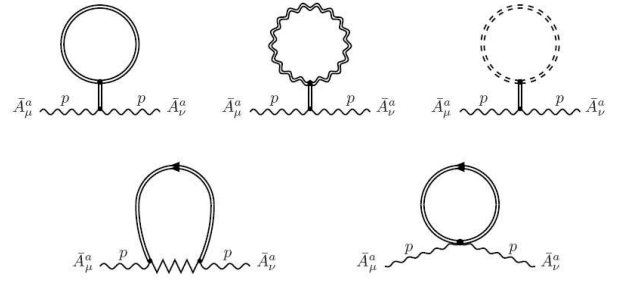}
\caption{\small The one-loop gravitational correction to the
two-point Green's function of the gauge field. The internal double,
double wavy and double dashed lines represent the gravitons
$h^{\vec{n}}_{\mu\nu}$, $A^{\vec{n}}_{i\mu}$ and
$\phi^{\vec{n}}_{ij}$ respectively. The internal zigzag lines
represent the quantum gauge fields $a^a_{\mu}$. The external wavy
lines are background gauge fields
$\bar{A}^a_{\mu}$.}\label{diagram1}
\end{center}
\end{figure}

Now we are going to calculate the one-loop gravitational correction
in the DR scheme. In the present setup, we have the relevant
interaction terms $h^{\vec{n}}_{\mu\nu}a_{\rho}\bar{A}_{\sigma}$,
$h^{\vec{n}}_{\mu\nu}\bar{A}_{\rho}\bar{A}_{\sigma}$,
$h^{\vec{n}}_{\mu\nu}h^{\vec{m}}_{\rho\sigma}h^{\vec{l}}_{\lambda\tau}$,
$h^{\vec{n}}_{\mu\nu}A^{\vec{m}}_{i\rho}A^{\vec{l}}_{j\sigma}$ and
$h^{\vec{n}}_{\mu\nu}\phi^{\vec{m}}_{ij}\phi^{\vec{l}}_{kl}$ at
$\kappa$ order, and
$h^{\vec{n}}_{\mu\nu}h^{\vec{m}}_{\rho\sigma}\bar{A}_{\lambda}\bar{A}_{\tau}$
at $\kappa^2$ order. The Feynman diagrams are shown in
Fig.\ref{diagram1} up to $\kappa^2$ order. A tedious calculation
shows that all of the first three diagrams are zero, though they are
not necessary vanished to all appearances because the Kaluza-Klein
gravitons are massive.

The last two diagrams have nontrivial contribution to the $\beta$
function. By a straightforward calculation, it can be shown that the
correction obtained in the framework of background field method is
independent of the gauge parameter $\xi$, as it should be. The
result of the diagram is
\begin{eqnarray}
\Pi_{\mu\nu}&=&i\frac{1}{\varepsilon}\frac{\kappa^2}{24\pi^2} (p^2
\eta_{\mu\nu}- p_{\mu} p_{\nu})\sum_{\vec{n}}(p^2
+18m^2_{\vec{n}})\nonumber\\
&&+[finite\;\; part]. \label{resultg}
\end{eqnarray}
The contributions of the extra dimensions are embodied in the
summation over the Kaluza-Klein states in a tower. In the limit of
four-dimensional spacetime, i.e. n=0, the second term is vanished
because the graviton is massless.

As discussed in Refs. \cite{wu1,wu2}, the first term of Eq.
(\ref{resultg}) corresponds to a higher derivative operator with
mass dimension six $-{1\over 2}
\partial _{\mu}\bar{F}^{a\mu\nu}
\partial^{\rho}\bar{F}^a_{\rho\nu}$. It can generate a Lee-Wick gauge field. The second term results from the summation of the
active Kaluza-Klein states. It contributes to the one-loop
$\beta$-function of the gauge coupling
\begin{equation}
\beta=-\frac{b}{(4\pi)^2}g^3+\frac{12g\kappa^2}{(4\pi)^2}\sum_{\vec{n}}m_{\vec{n}}^2,
\label{betaf}
\end{equation}
where $b$ depends on the gauge and matter contents and is wholly
independent of whether gravitation is included in the calculation.
It comes from the first term of Eq.(\ref{beta1}). The second term in
Eq.(\ref{betaf}) is contributed by the exchange of graviton
Kaluza-Klein states, and is universal for all gauge couplings since
gravitons do not carry any gauge charge. It tends to make the theory
anti-asymptotically free.

In the theory where $b$ is positive, such as QCD and the SU(2) gauge
sector of the electroweak, there is a nontrivial ultraviolet-stable
fixed point at \begin{eqnarray}
g_{\ast}^2&=&\frac{12\kappa^2}{b}\sum_{\vec{n}}m_{\vec{n}}^2.\label{uv}
\end{eqnarray}
In the limit of large momentum, the coupling constant tends to
$g_{\ast}$, and therefore runs along with $\kappa$.

  The summation can be written as an integration
in term of the mass $m^2_{\vec{n}}$ when the Kaluza-Klein states are
near degenerate \cite{Han}
\begin{equation}
\sum_{\vec{n}}f(m_{\vec{n}})=\int_0^{\Lambda^2}
dm^2_{\vec{n}}\rho(m_{\vec{n}})f(m_{\vec{n}}),
\end{equation}
where
\begin{equation}
\rho(m_{\vec{n}})=\frac{R^n
m^{n-2}_{\vec{n}}}{(4\pi)^{n/2}\Gamma(n/2)}
\end{equation}
is the Kaluza-Klein state density. As is well-known, using DR
scheme, which is a mass-independent scheme, heavy states do not
decouple. Thus we have introduced an explicit cutoff $\Lambda$ to
regularize the mass integration. That is, we include only a finite
number of low-lying Kaluza-Klein states whose masses are smaller
than $\Lambda$ and assume all other states decouple from the low
energy physics we are interested. This cutoff does not break any
gauge symmetry and our calculation is gauge independent. The near
degenerate condition is satisfied when the energy scale $R^{-1}$
characterized the first Kaluza-Klein excitations is much less than
the physical scale $\Lambda$.

In the literatures, there are two different scenarios for
calculating the $\beta$ function from the Kaluza-Klein states,
depending on how to treat the cutoff scale $\Lambda$ and the
evolution scale $\mu$.

One is that the energy scales $\Lambda$ and $\mu$ are different,
which implies that, when $\mu<\Lambda$, the particles with mass
larger than $\mu$ can contribute to the gauge coupling. This is the
case for the QCD where the heavy quark flavors are of significance
to the running of gauge coupling at low energy. In general, the
cutoff scale $\Lambda$ should be less than $M_{pl(4+n)}$, since the
effective theory is only expected to be valid below the fundamental
scale $M_{pl(4+n)}$. Unitarity requirement of some explicit extra
dimensional models can play some stringent constraints on it
\cite{he}.

The gauge coupling runs logarithmically
\begin{equation}
g(\mu)^2=\frac{g(E)^2}{1+\frac{2b}{(4\pi)^2}g(E)^2\ln{\mu\over
E}-a(\frac{\Lambda}{M_{pl(4+n)}})^{n+2}\ln{\mu\over E}},
\label{coupling}
\end{equation}
where the coefficient $a$ is purely determined by the number of
extra dimensions \begin{equation}
a=\frac{48}{(n+2)\pi(4\pi)^{\frac{n}{2}}\Gamma(\frac{n}{2})}.
\end{equation}
The $\Gamma({n\over 2})^{-1}$ factor ensures that the gravitational
correction is vanished in the limit of four spacetime dimension,
which is in agreement with the result in
\cite{Pietrykowski,wu1,toms,ebert}.

An interesting possibility follows from the UV fixed point. When the
Kaluza-Klein cutoff scales $\Lambda_3$ and $\Lambda_1=\Lambda_2$ in
the SM are elaborately set to \begin{eqnarray}
\frac{\Lambda_2^{n+2}}{b_{2}}=\frac{\Lambda_3^{n+2}}{b_{3}}=\frac{(n+2)(4\pi)^{n/2}\Gamma(n/2)\mu^4\kappa^2}{24(16\pi)^2R^{n}},
\end{eqnarray}
the non-abelian gauge coupling constants $g_2$ and $g_3$ and that of
the gravity $g_{\kappa}$ unify at the UV-stable fixed points
$g_{2\ast}=g_{3\ast}=g_{\kappa}$, where $g_{\kappa}$ is the
dimensionless Newton constant, and $g_{\kappa}\equiv
\mu^2\kappa^2/16\pi$ \cite{lauscher}. If $g_1$ can evolve to the
fixed point at some high energy scale, the four coupling constants
is unified.

The other scenario that the energy scales $\Lambda$ and $\mu$ are
the same will lead to the power law running of the gauge coupling
\begin{equation}
g(\mu)^2=\frac{g(E)^2}{1+\frac{2b}{(4\pi)^2}g(E)^2\ln{\mu\over
E}-a\frac{\mu^{n+2}-E^{n+2}}{M_{pl(4+n)}^{n+2}}}, \label{coupling}
\end{equation}
with
\begin{equation} a=\frac{24}{(n+2)\pi(4\pi)^{{n\over
2}}\Gamma({n\over 2})}.
\end{equation}
It implies that only the Kaluza-Klein gravitons with mass less than
$\mu$ make contributions to the running of the gauge couplings. Note
that the one-loop correction to the gauge couplings from a tower of
gauge boson Kaluza-Klein states is of order $\mu^{n}$ \cite{Dienes}.
Here we show explicitly that the correction coming from graviton
Kaluza-Klein states is of order $(\mu/M_{pl(4+n)})^{n+2}$.

The research of F.W. is supported in part by the project of Chinese
Ministry of Education (No. 208072) and the Program for Innovative
Research Team of Nanchang University. M.Z. is supported in part by
the research fund of National University of Defense Technology.

\end{document}